\documentclass[5p,authoryear,12pt]{elsarticle}
\usepackage{lineno} 
\usepackage{color,upgreek, microtype, amsfonts, amsmath} 
\usepackage[ colorlinks=true,%
]{hyperref}

\modulolinenumbers[5]

\journal{Journal of Atmos. and Sol-Terr. Phys.}

\graphicspath{{/Users/erdal/Research/Storm/Figures/Final/}}

\begin{document}
\begin{frontmatter}

\title{Hemispheric Differences in the Response of the Upper Atmosphere to the
  August 2011 Geomagnetic Storm: A Simulation Study}

\author[gmu]{Erdal Yi\u git\corref{cor}}
\cortext[cor]{Corresponding author}
\ead{eyigit@gmu.edu}

\address[gmu]{Space Weather Laboratory, Department of Physics and Astronomy,
  George Mason University, Fairfax, Virginia, USA}

\author[ssl]{\rm Harald U. Frey}
\address[ssl]{Space Sciences Laboratory, University of California at
  Berkeley, Berkeley, USA}

\author[mich]{\rm Mark B. Moldwin}
\address[mich]{Atmospheric Oceanic and Space Sciences, University of Michigan,
  Ann Arbor, USA}

\author[ssl]{\rm Thomas J. Immel}
\author[mich]{\rm Aaron J. Ridley}

\begin{abstract}
  Using a three-dimensional nonhydrostatic general circulation model, we
  investigate the response of the thermosphere-ionosphere system to the 5--6
  August 2011 major geomagnetic storm. The model is driven by measured
  storm-time input data of the Interplanetary Magnetic Field (IMF), solar
  activity, and auroral activity.  Simulations for quiet steady conditions over
  the same period are performed as well in order to assess the response of the
  neutral and plasma parameters to the storm. During the storm, the
  high-latitude mean ion flows are enhanced by up to 150--180\%. Largest ion
  flows are found in the main phase of the storm. Overall, the global mean
  neutral temperature increases by up to 15\%, while the maximum thermal
  response is higher in the winter Southern Hemisphere at high-latitudes than
  the summer Northern Hemisphere: 40\% vs. 20\% increase in high-latitude mean
  temperature, respectively. The global mean Joule heating increases by more
  than a factor of three. There are distinct hemispheric differences in the
  magnitude and morphology of the horizontal ion flows and thermospheric flows
  during the different phases of the storm. The largest hemispheric difference
  in the thermospheric circulation is found during the main and recovery phases
  of the storm, demonstrating appreciable geographical variations.The advective
  forcing is found to contribute to the modeled hemispheric differences.

\end{abstract}


\end{frontmatter}

  \section{Introduction}
  The thermosphere ($\sim 90-500$ km) is the upper part of the neutral
  atmosphere that coexists with and is dynamically and chemically coupled to the
  ionosphere via electrodynamical and wave processes. The thermosphere is
  influenced by lower atmospheric internal waves \citep{Altadill_etal04,
    Abdu_etal06, Pancheva_etal09, YigitMedvedev10, YigitMedvedev15} and exhibits
  large solar and geomagnetic variations \citep{Emery_etal99, Immel_etal01,
    Balan_etal11}. Owing to the simultaneous interplay of upward wave
  propagation, space weather effects, and internal processes, the upper
  atmosphere has substantial spatiotemporal variability \citep{InnisConde02a,
    Matsuo_etal03, Bristow08, Anderson_etal11, Kil_etal11, PanchevaMukhtarov11,
    Yigit_etal12a}.

  Geomagnetic storms are a central aspect of ``space weather" that can have a
  great impact on, for example, airline crew and passengers, telecommunication
  systems, electric power grids, and satellite navigation. Therefore space
  weather is considered a subject of natural hazard. The origin of space weather
  is the high-energy plasma ejection by the Sun into the interplanetary space.
  Sudden high-density plasma ejection processes can be detected by the
  NASA-European Space Agency (ESA) Solar and Heliospheric Observatory (SOHO)
  spacecraft. An extreme form of plasma ejection is the coronal mass ejection
  (CME) that is a large eruption of magnetic field and plasma from the outer
  atmosphere of the Sun. A CME directed toward Earth with a typical speed of
  1000 km s$^{-1}$ would take about 40 hours to reach Earth and can produce
  major geomagnetic storms.  Improved predictions though are necessary to be
  able to take the necessary precautions in cases of extreme storm events
  However, space weather predictions are relatively limited. Even in the case of
  terrestrial lower atmospheric weather, predictions can be done only to a
  certain extent, despite the advances in modern technology that allows
  a detailed monitoring.
  
  In broader terms, space weather is understood as the effects of the
  Sun-magnetosphere system on Earth's atmosphere-ionosphere system.  What is
  then a geomagnetic storm? \citet{Gonzalez_etal94} describe a geomagnetic
  storm as a period during which sufficiently intense long-lasting
  interplanetary convection electric fields are present and the
  magnetosphere-ionosphere system is substantially energized. That is, an
  interplanetary electric field causes convection in the
  magnetosphere-ionosphere system.

  Geomagnetic storms can have great impact on Earth's environment. Storm-induced
  thermospheric mass density enhancement can affect satellite orbital lifetimes
  and expectancy \citep{Prolss11} and can make Global Position System signals
  undergo rapid fluctuations in amplitude and phase. While terrestrial
  ``meteorological storms" (e.g., hurricanes, thunderstorms) have direct impact
  on Earth's lower atmosphere, the influences of ``space weather storms" on
  Earth's atmosphere are indirect, primarily via the changes in the
  magnetosphere and the subsequent downward coupling to the
  atmosphere-ionosphere system therefrom. The intensity of geomagnetic storms
  are extremely variable \citep{Mannucci_etal05}.

  Historically, magnetic storms were discussed theoretically before the advent
  of global satellite technology \citep[e.g.,][]{Martyn51}. Later, computer
  models have been developed based on theoretical equations in order to
  predict the response of the upper atmosphere to storms.  Earlier numerical
  modeling efforts have provided valuable insight into high-latitude ionospheric
  convection and the effects of geomagnetic storms. With a longitudinally
  invariant hydrostatic numerical model that assumes a coincident geomagnetic
  field with the Earth's rotational axis, \citet{RichmondMatsushita75}
  demonstrated equatorward propagating large-scale gravity waves (GWs) during a
  substorm. Equatorward propagating large-scale GWs can transfer energy and
  momentum from the high-latitudes to low-latitudes. General circulation
  modeling efforts by \citet{Roble_etal82} excluding the effects of particle
  precipitation showed that ionospheric plasma convection constitutes a
  substantial amount of momentum and energy source at high-latitudes.
  
  Various aspects of geomagnetic storms have been investigated by
  three-dimensional nonlinear hydrostatic upper atmosphere models
  \citep[e.g.,][]{Fuller-RowellRees81, Emery_etal99, Zaka_etal10, Burns_etal12,
    Lu_etal13} as well as by observations \citep[e.g.,][]{Oliver_etal88,
    Anderson_etal98, Balan_etal11, Balan_etal12, Kil_etal11, Ma_etal12}. Today,
  advances in satellite technologies enable continuous monitoring and
  characterization of the geospace environment and computer models are used to
  diagnose the effects of solar and geomagnetic variations on Earth's upper
  atmosphere.

  Recently, a series of geomagnetic storm events occurred between
  August--October 2011 that motivated researchers to better characterize these
  storms \citep[e.g.,][]{Earle_etal13, Gong_etal13, Haaser_etal13,
    Blanch_etal13, Huang_etal14}. While increased amount of observations of the
  geospace during storms help researchers characterize these storms in more
  detail, application of coupled nonlinear global models can provide a framework
  to determine possible storm effects on the upper atmosphere. Satellite
  observations can be used to drive global atmosphere-ionosphere models that
  require boundary conditions with measured magnetospheric variations in order
  to simulate the consequences of storms on the upper atmosphere, i.e., the
  thermosphere-ionosphere.

\begin{figure*}[t]
\centering
  \includegraphics[width=0.7\textwidth]{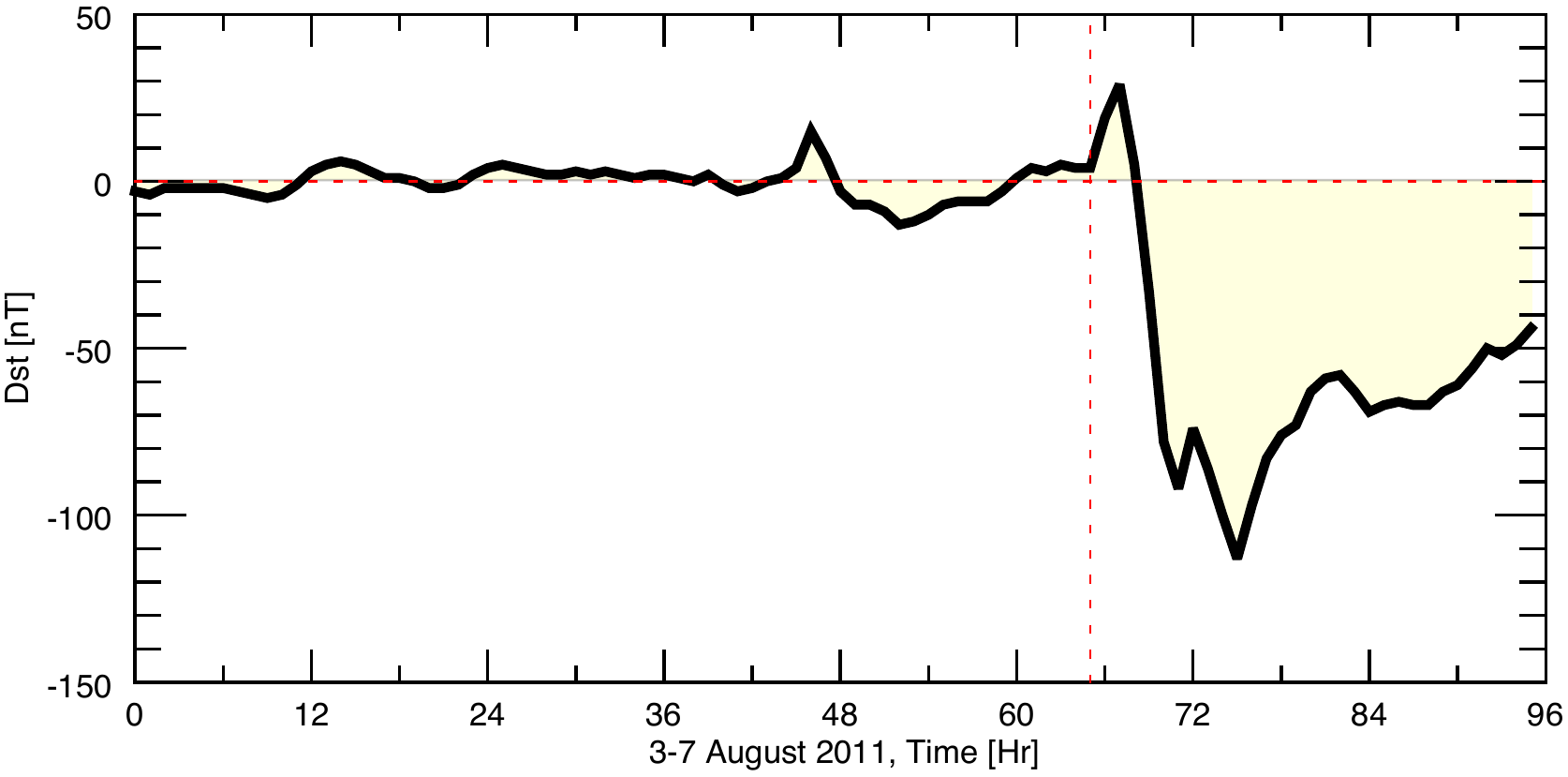}
  \caption{Variations of the Disturbed Storm Time index ($Dst$) retrieved from
    the World Data Center for Geomagnetism, Kyoto, in nT from 3 to 7 August 2011
    (96-hour period). The horizontal and
    vertical dotted red lines denote the 0 nT level and
    the beginning of the disturbed period, respectively.}
  \label{fig:dst}
\end{figure*}

  The goal of this paper is to apply general circulation modeling techniques to
  investigate the storm-time global changes in the upper atmosphere in ways that
  have not done before, to address the previously challenging problem of
  hemispheric differences and to understand the related dynamical
  mechanisms. The unique aspects of the August 2011 geomagnetic storm offer an
  ideal case to extend previous studies, and having done so we can offer new
  physical insights in geomagnetic storm forcing and the resultant effects in
  the upper atmosphere. Specifically, we use the Global Ionosphere Thermosphere
  Model \citep[(GITM),][]{Ridley_etal06}, which is a three-dimensional nonlinear
  nonhydrostatic general circulation model (GCM) that self-consistently couples
  the thermosphere and ionosphere. Observations of the August 2011 geomagnetic
  storm from NASA's Advanced Composition Explorer (ACE) satellite will be used
  as model input in order to perform simulations under realistic storm
  conditions. The main focus of the paper is the dynamical response of the upper
  atmosphere to the storm.

  The structure of the paper is as follows: The next section describes the
  observations of the August 2011 geomagnetic storm; sections
  \ref{sec:model}--\ref{sec:exp} introduce the GITM model and the configurations
  of model simulations. Simulation results of the ionospheric high-altitude
  convection patterns and ion speeds are presented in section
  \ref{sec:conv}. The dynamical and thermal responses of the thermosphere to the
  storm are investigated in section \ref{sec:thermo}. Section \ref{sec:asym}
  presents the storm-induced hemispheric differences in the
  thermosphere. Finally, a discussion of the results are provided in section
  \ref{sec:discussion} and conclusions are drawn in section \ref{sec:conc}.

  %
  %
  \section{August 2011 Geomagnetic Storm}
  \label{sec:storm}

\begin{figure*}[!t]
\centering
  \includegraphics[width=0.75\textwidth]{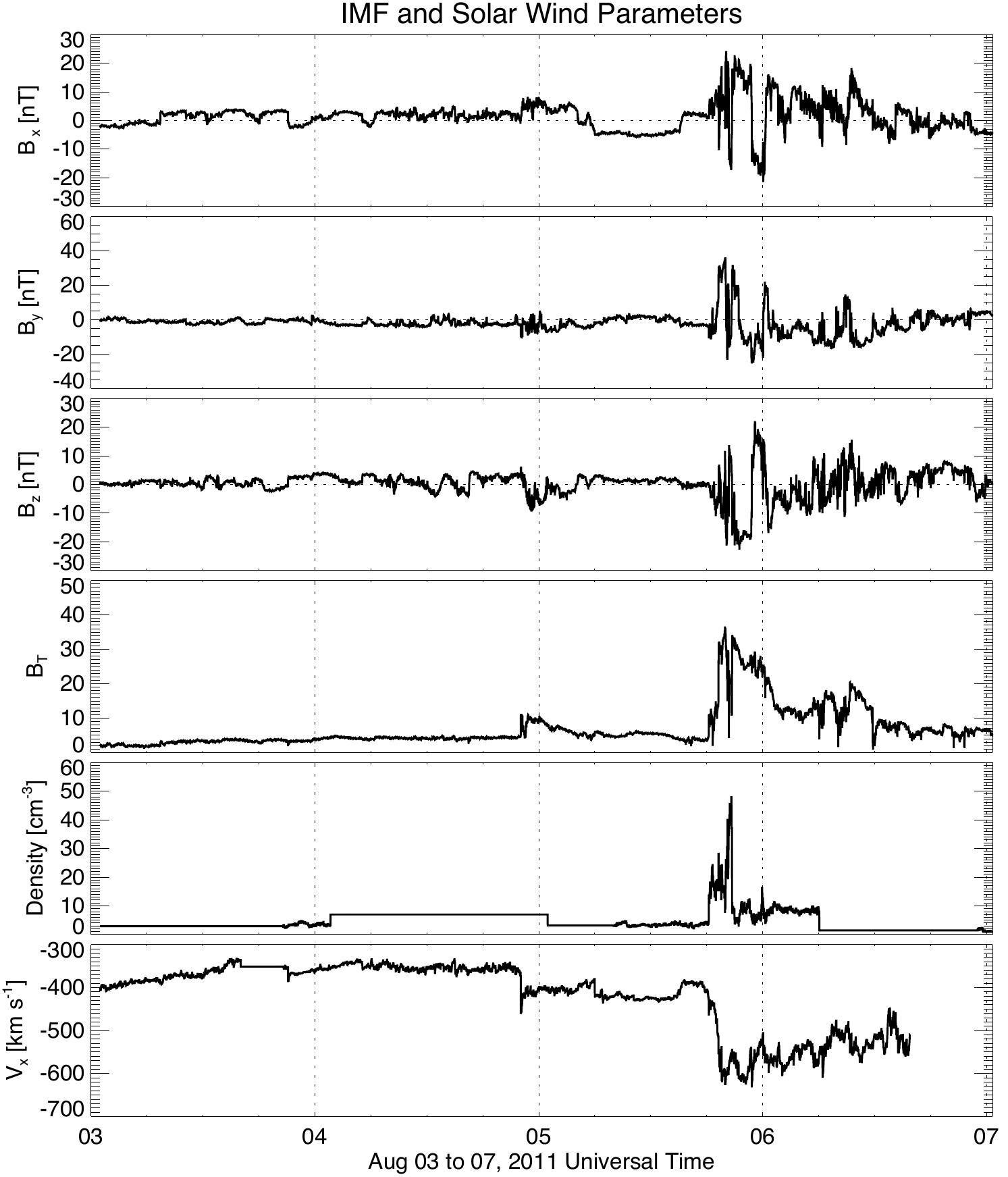}
  \caption{Variations of the interplanetary magnetic field (IMF) components
    $B_x$, $B_y$, $B_z$, the total IMF magnitude $B$, solar wind density and
    speed $v_x$ from 3 to 7 August 2011 observed by NASA's Advanced
      Composition Explorer (ACE) satellite.}
  \label{fig:imf}
\end{figure*}

  Geomagnetic activity can be examined using various parameters, such as the
  interplanetary magnetic field (IMF) ${\bf B}_{IMF}$, $Dst$, $K_{p}$, and $AE$
  indices. The Disturbed Storm Time ($Dst$) index describes the change of the
  Earth's internal magnetic field from its standard quiet time value, therefore
  a negative $Dst$ indicates a decrease in Earth's magnetic field and is due to
  the increase in the magnetospheric ring currents. Figure \ref{fig:dst} shows
  the variation of the $Dst$ index in nT from 3 August 0000 UT to 7 August 0000
  UT 2011 obtained from the World Data Center for Geomagnetism, Kyoto. During
  quiet (undisturbed) times, the magnitude of this disturbance field is
  small. Around 1800 UT on 5 August $Dst$ starts becoming negative, reaching a
  minimum of 110 nT around 0300 UT on 6 August, indicating the occurrence of a
  moderate geomagnetic storm. After 0300 UT, we observe the recovery phase of
  the storm indicated by $Dst$ gradually becoming less negative.

  The IMF is an extension of the solar magnetic field into the heliosphere and
  connects to Earth's intrinsic magnetic field, i.e., the geomagnetic field,
  allowing high-energy particles to reach the Earth. Strictly speaking, the IMF
  is a driver and the other indices describe the resultant geomagnetic activity.
  Perturbations and enhancements in the IMF are therefore a way for various
  solar effects to ``communicate" with the geospace. Figure \ref{fig:imf}
  presents the temporal variations of the interplanetary magnetic field
  components, $B_x$, $B_y$, and $B_z$, the magnitude $B=|{\bf B}_{IMF}|$ of
  the IMF, the solar wind density and speed $v_x$ from 3 to 7 August 2011
  observed by NASA's Advanced Composition Explorer (ACE) satellite. Launched in
  1997, the ACE satellite measures the characteristics of the interplanetary
  medium. Magnetic field measurements are performed by the MAG (Magnetic Field
  Experiment) instrument while the solar wind is analyzed by SWEPAM (Solar Wind
  Electron Proton Alpha Monitor) on board ACE. These data are collected at high
  time resolution of 15-25 s.

  At around 1800 UT on 5 August a major geomagnetic storm commences (initial
  phase), due to the compression of the magnetosphere by the arrival of a
  high-density solar wind that is seen in Figure \ref{fig:imf}. An increase in
  the solar wind speed is seen as well. Associated with this process, all ${\bf
    B}_{IMF}$ components undergo rapid large fluctuations and the total IMF
  magnitude increases from its 0--5 nT quiet values up to 40 nT within few
  hours. Enhancement of the southward IMF, coincident with the negative $Dst$
  (Figure \ref{fig:dst}), indicates the main phase of the storm starting around
  2100--2200 UT. The IMF then gradually decreases after 0300 UT on 6 August
  indicating the recovery phase of the storm. The main phase of the storm lasts
  for about five hours and the southward IMF reaches peak values of --20 nT,
  which is often characteristic of a major geomagnetic storm. The southward
  component of the IMF is an important parameter because it is a
  proxy for an interconnection between the geomagnetic and the
  interplanetary magnetic field lines \citep{Dungey61}. Overall, remarkable
  temporal variations are seen in all IMF components and the solar wind. Next,
  we will describe GITM, which we will use to investigate the impact of this
  storm on the thermosphere-ionosphere system.

  %
  %
  \section{Model Description}
  \label{sec:model}
  GITM is a three-dimensional first principle nonlinear nonhydrostatic
  time-dependent General Circulation Model (GCM) extending from 100 km to the
  thermosphere at $\sim$ 600 km. It solves the equations of momentum, continuity
  and heat transport for neutrals and ions self-consistently. The plasma and
  neutral equations are coupled, that is, they are solved simultaneously for
  charged and neutral species. The model is described fully in the work of
  \citet{Ridley_etal06} and has been used a number of times to investigate
  thermosphere-ionosphere coupling processes
  \citep[e.g.,][]{PawlowskiRidley09,YigitRidley11a, YigitRidley11b,
    Yigit_etal12a}. Momentum, continuity, and energy equations are iteratively
  solved with a time step of 2--4 s, which allows the simulation of highly
  variable upper atmosphere processes. In the simulations to be presented, a
  latitude-longitude grid of $2.5^\circ\times5^\circ$ is used. This resolution
  yields a latitude grid spacing of $dy = 211.3$ km, minimum and maximum
  longitude grid spacing of 12.1 km and 553.3 km, respectively.  In the vertical
  direction, the resolution is one third scale height with 54 altitude
  levels. The model can be run for varying conditions of solar and geomagnetic
  activity, using fixed values as well as observed solar fluxes and geomagnetic
  parameters. The electrodynamics is self-consistently solved as described in
  the work by \citet{Vichare_etal12}, including the dynamo electric
  field. During storm-time the associated disturbance electric fields appear in
  the equatorial region \citep{KlimenkoKlimenko12}. The high-latitude electric
  potential patterns are prescribed by the empirical model of \citet{Weimer96}
  and the particle precipitation is after the work by
  \citet{Fuller-RowellEvans87}. The Weimer model uses the $B_y$ and $B_z$
  components of the IMF and tilt angle of the dipole with respect to the
  ecliptic as input. The particle precipitation model uses the hemispheric power
  index as input. These high-latitude empirical models do not contain realistic
  short-time variability; they are average models.

  At the lower boundary diurnal and semidiurnal migrating and nonmigrating solar
  tidal fields are obtained from the Global Scale Wave Model (GSWM) of
  \citet{HaganForbes02, HaganForbes03}.
  The vertical momentum equation is solved explicitly, allowing the model to
  simulate nonhydrostatic effects and acoustic gravity waves
  \citep{Deng_etal08}. General circulation studies with
  GITM conducted by  \citet{YigitRidley11b} and \citet{Yigit_etal12a}
  demonstrated the importance of nonhydrostatic acceleration for thermospheric
  dynamics even during relatively quiescent geomagnetic and solar conditions. 
  %
  %
  \section{Experiment Design and Model Simulations}
  \label{sec:exp}

\begin{figure*}[!t]\centering
  \includegraphics[width=0.60\textwidth]{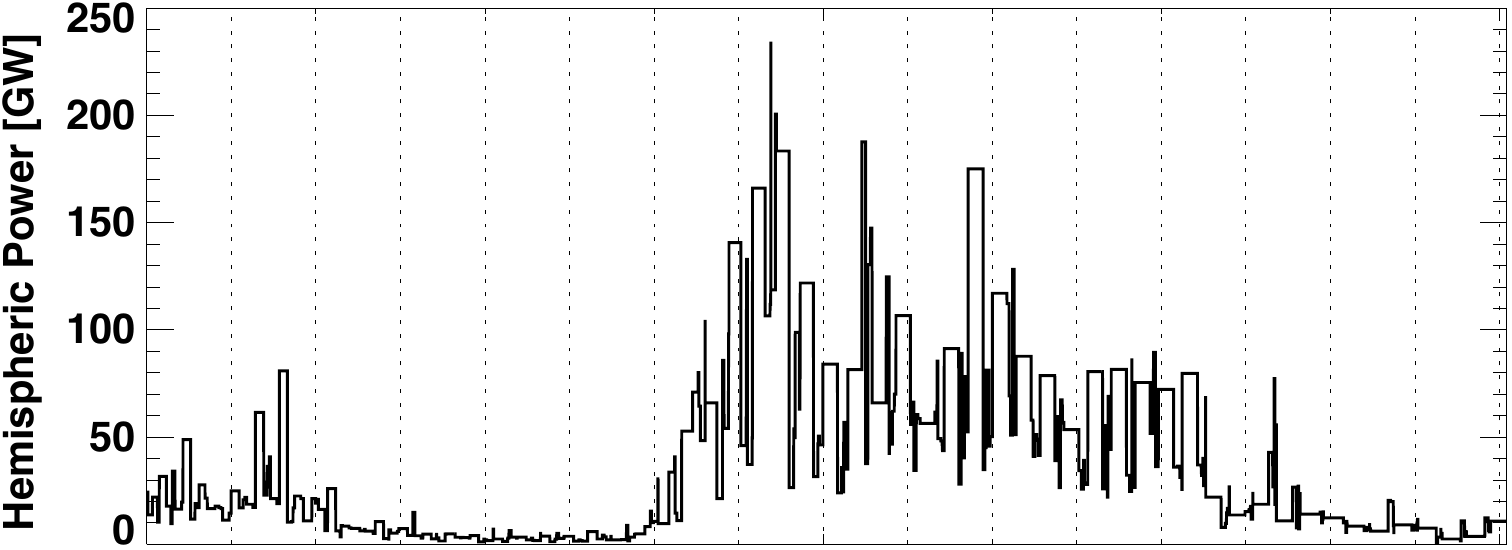}
  \includegraphics[width=0.60\textwidth]{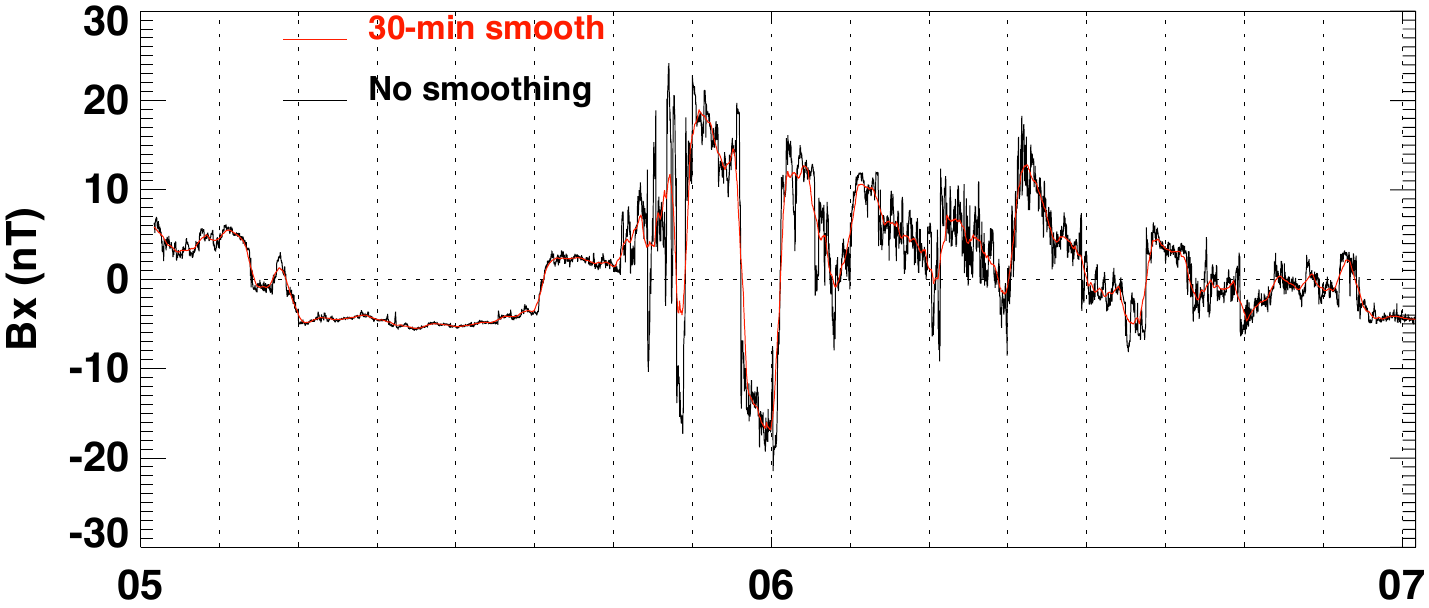}
  \includegraphics[width=0.60\textwidth]{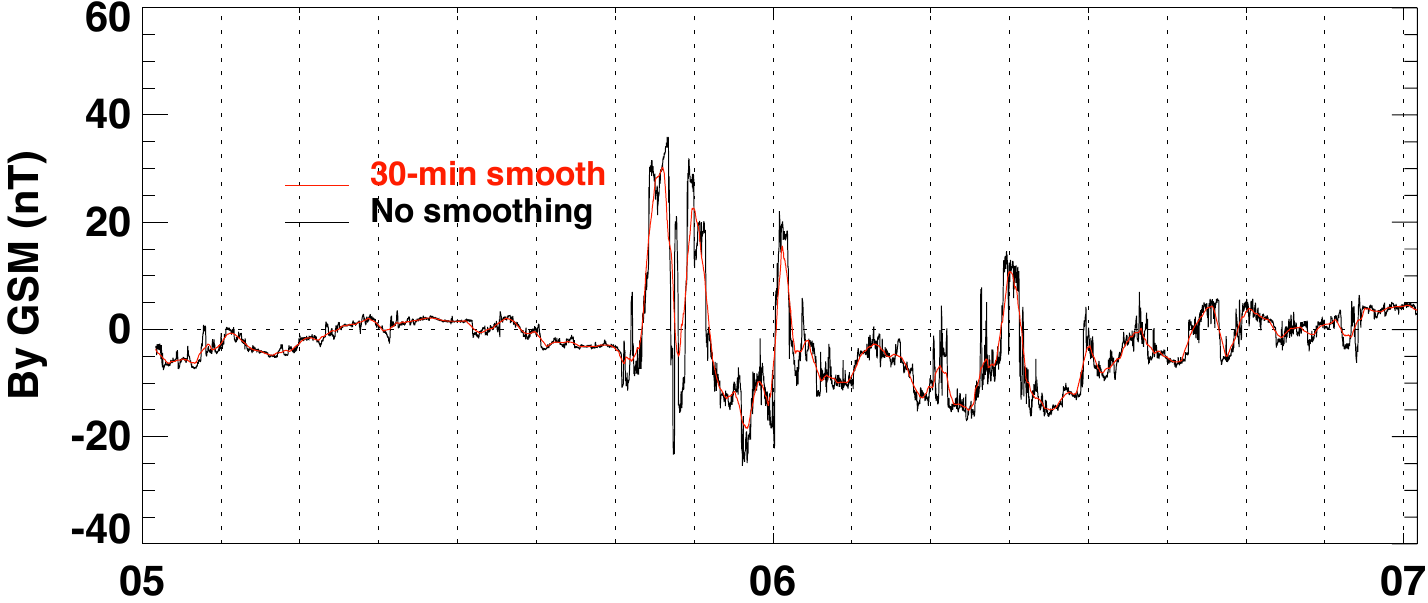}
  \includegraphics[width=0.60\textwidth]{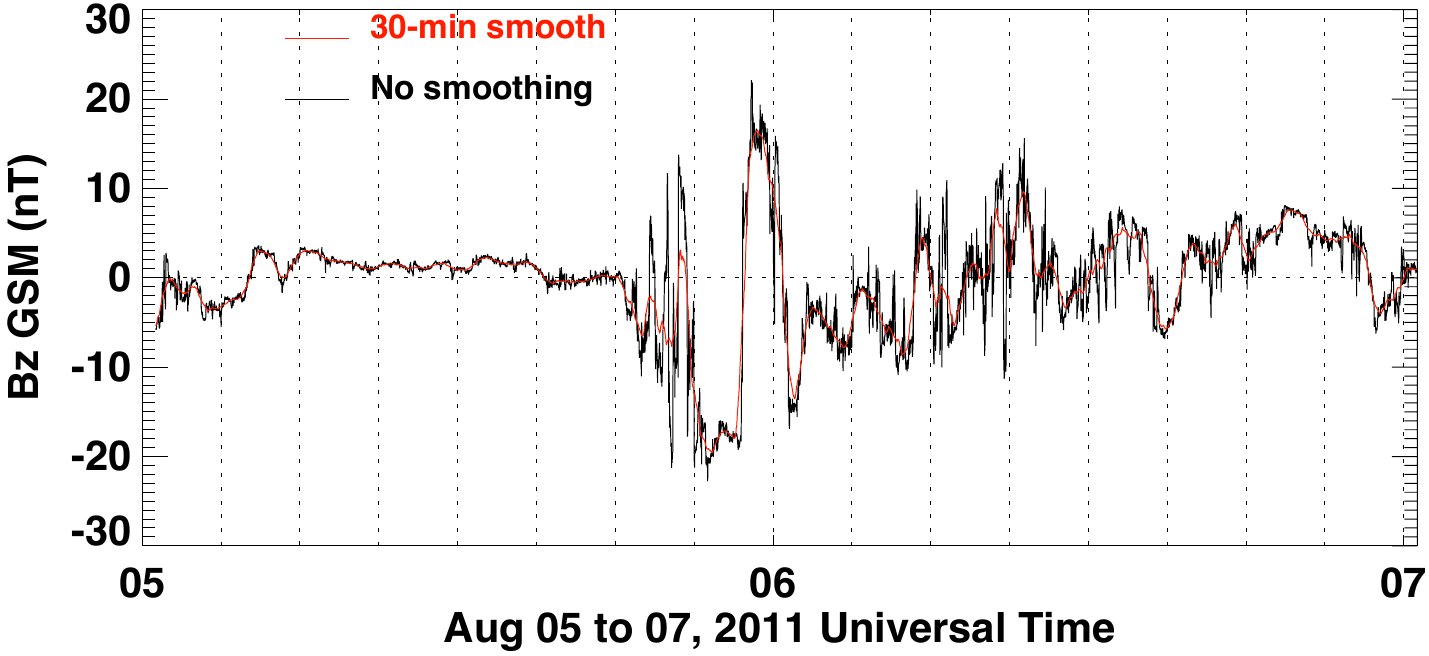}
  \caption{Observed values (black) of the hemispheric power in $10^9$ W (one
    giga Watt) and of
    the IMF components $B_x$, $B_y$, and $B_z$ and their 30-min smoothed
    variations (red) from 5 to 7 August 2011.}
  \label{fig:input}
\end{figure*}

  Figure \ref{fig:imf} shows that the components of ${\bf B}_{IMF}$ and the
  solar wind are highly variable during the August storm. The solar wind
  influences the magnetosphere which in turn affects the
  thermosphere-ionosphere. These processes modify the upper atmosphere in
  addition to the effects of the direct solar input. To assess the response of
  the thermosphere in the presence of highly variable forcing can be a
  challenging task. Models provide the capability of conducting controlled
  simulations to diagnose the effects of different dynamical processes on the
  system. 

  We first conduct a ``benchmark" simulation based on the average values
  during August 1--5 of the observed geomagnetic parameters with hemispheric
  power (HP) of $3.68$ GW, $B_x = 1.42$ nT, $B_y = -1.57$ nT, $B_z = -0.85$ nT
  and solar wind speed of $v_x \approx 363$ km s$^{-1}$. GITM is run with these
  constant values from 1 to 5 August 0000 UT and the last output (5 August 0000
  UT) is saved as a starting point (start-up time) for the subsequent
  simulations. Then, two model simulations have been conducted from 5 to 7
  August 0000 UT, covering the storm period. First, with the configuration of
  the steady (benchmark) run and second, 30-minute smoothed observed IMF
  data are used as inputs as shown in Figure \ref{fig:input} with red lines,
  while the black line represents the actual observations. This smoothing
  enables us to focus on the larger-scale variations of the magnetosphere and
  subsequently determine the overall response of the thermosphere to the
  geomagnetic storm rather than to the small-scale temporal variations that are
  seen in observations. While there is large temporal variability in the actual
  observations, smoothing gives a general picture of how geomagnetic parameters
  vary.  Small-scale temporal variability during storms is the subject of our
  future modeling efforts.


  %
  %
  \section{High-Latitude Ionospheric Convection}
  \label{sec:conv}

  During geomagnetic storms, the collisionless solar wind with velocity ${\bf
    v}_{sw}$ distorts the Earth's magnetic field significantly.  The solar wind
  induced magnetospheric electric field ${\bf E}$ in Earth's frame of
  reference is
  \begin{linenomath*}
    \begin{equation}
      {\bf E} = -{\bf v}_{sw}\times {\bf B},
      \label{eq:efield}
    \end{equation}
  \end{linenomath*}
  which is mapped down to the polar cap ionosphere. An increase in the solar
  wind speed and/or magnetic field ${\bf B}$ leads to an increase in ${\bf E}$.
  Electric fields originating from the magnetosphere make the plasma ``convect"
  at high-latitudes and can accelerate the ions to relatively large speeds of
  more than 1 km s$^{-1}$ \citep{Crowley_etal89}. An accurate representation of
  ionospheric flow patterns largely driven by the magnetosphere is crucial for
  the investigation of the response of the thermosphere-ionosphere system to
  geomagnetic storms.  The convection patterns depend highly on the IMF
  direction and magnitude \citep{Bekerat_etal03}. While a southward directed IMF
  impacts plasma convection greatly in polar latitudes, a non-zero IMF $B_y$
  generated asymmetric thermospheric response to geomagnetic activity
  \citep{Yamazaki_etal15}.

  \subsection{High-Latitude Mean Ion Flows}

\begin{figure}[!t]\centering
  \includegraphics[width=\columnwidth]{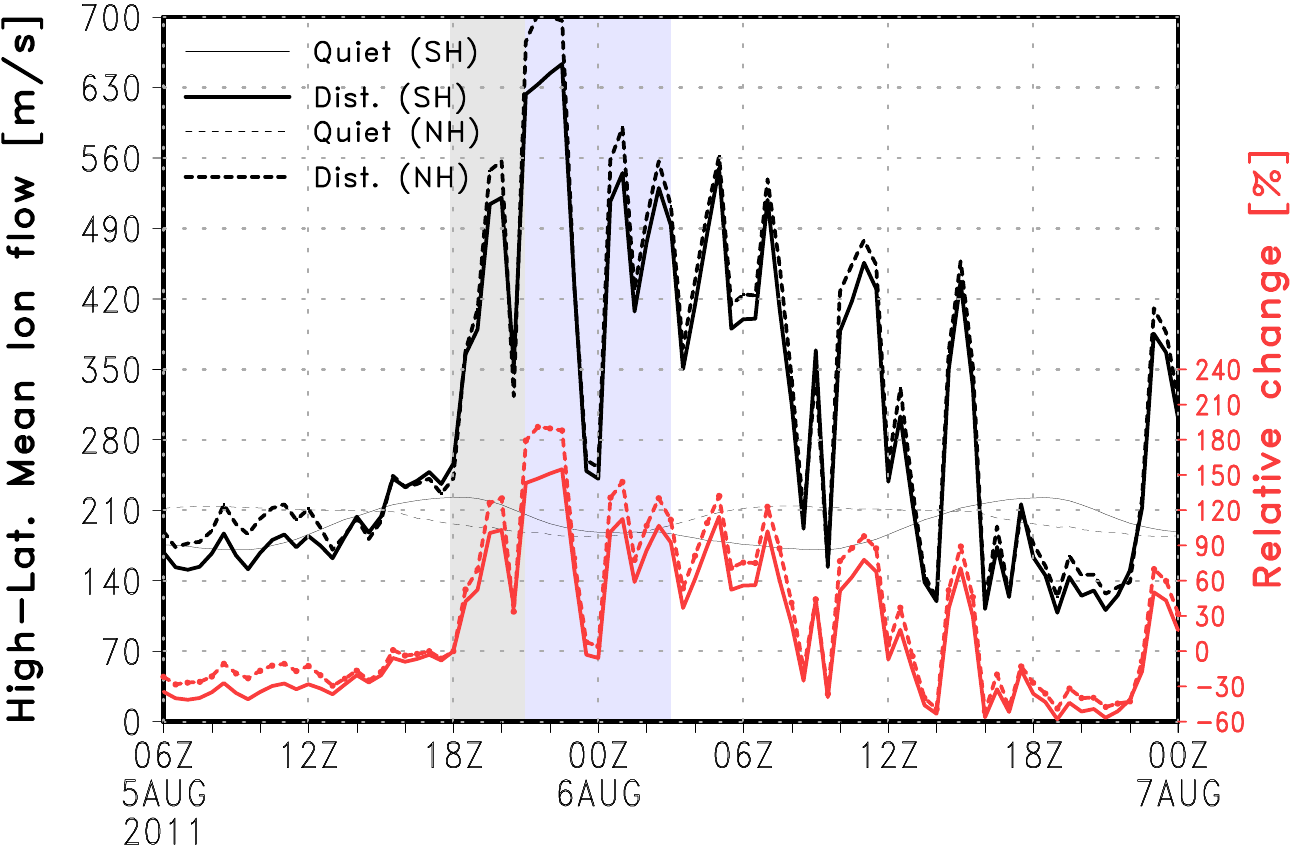}
  \caption{High-latitude means of horizontal ion flow at 400 km from 0600 UT 5
    August to 0000 UT 7 August. Benchmark and storm simulations are represented
    by thin and thick lines. Solid and dashed
    lines denote Southern and Northern Hemisphere high-latitude means,
    respectively. Red lines show the relative percentage changes with respect to
    the storm onset time (1800 UT, 5 August.). Each tick mark on the time axis
    represent 2-hour intervals. The initial and the main phases of the storm are
    shaded with gray and blue, respectively.}
  \label{fig:conv}
\end{figure}

  Because the primary influence of the storm is expected to occur in the
  ionosphere, which coexists with the thermosphere, we first investigate the ion
  flow patterns in the high-latitude ionosphere. Figure \ref{fig:conv} shows how
  the high-latitude means of the horizontal ion flows $|{\bf v_{i_{\rm H}}}|$
  change at 400 km as a function of time from 5 August 0600 UT, that is,
  starting about 12 hours before the onset of the storm, to 7 August 0000
  UT. The gray and blue shadings mark the initial and the main phases of the
  storm, respectively. The high-latitude means that are to be presented in the
  rest of the paper are calculated taking into account data poleward of
  (geographic) 60$^\circ$N/S. Thin and thick lines represent ``quiet"
  (benchmark) and ``storm" (disturbed) runs, respectively, while solid and
  dashed lines show the Southern and Northern Hemispheres results,
  respectively. The red lines show the relative percentage change with respect
  to the storm onset (1800 UT, 5 August).

  Under quiet conditions, the mean ion flows in the Northern and Southern
  Hemispheres vary diurnally with similar magnitudes but are slightly
  time-shifted with respect to each other primarily because of the offset
  between the geographic and geomagnetic poles. In the disturbed run, before the
  storm onset $|{\bf v_{i_{\rm H}}}|$ has similar magnitude as the benchmark
  run. As the storm commences, the shift in the UT variation of the mean ion
  flows in the different hemispheres is subdued as the IMF variations are the
  dominant driver of the ion flows. The response of the ionosphere is immediate
  in both hemispheres. Within the first 2 hours of the storm, the mean $|{\bf
    v}_{\rm {i}_H}|$ increases from 240 m s$^{-1}$ to $\sim$ 500 m s$^{-1}$ and
  to 560 m s$^{-1}$ in the Southern and Northern Hemispheres, respectively,
  corresponding to $\sim$ 100 \% and $>120$\% relative increase. In the main
  phase of the storm at around 2200 UT 5 August where the southward IMF is at
  its maximum (Figure \ref{fig:input}), the mean $|{\bf v_{i_{\rm H}}}|$ reaches
  a maximum for the entire simulation period: to $\sim$630 m s$^{-1}$ and $>700$
  m s$^{-1}$ in the Southern and Northern Hemispheres, respectively,
  corresponding to $\sim 150 \%$ and $\sim 200 \%$ relative increases. The mean
  $|{\bf v_{i_{\rm H}}}|$ then gradually decreases in a similar rate in both
  hemispheres but remains generally large during the main phase and the recovery
  phase of the storm.

\begin{figure}[!t]\centering 
  \includegraphics[width=\columnwidth]{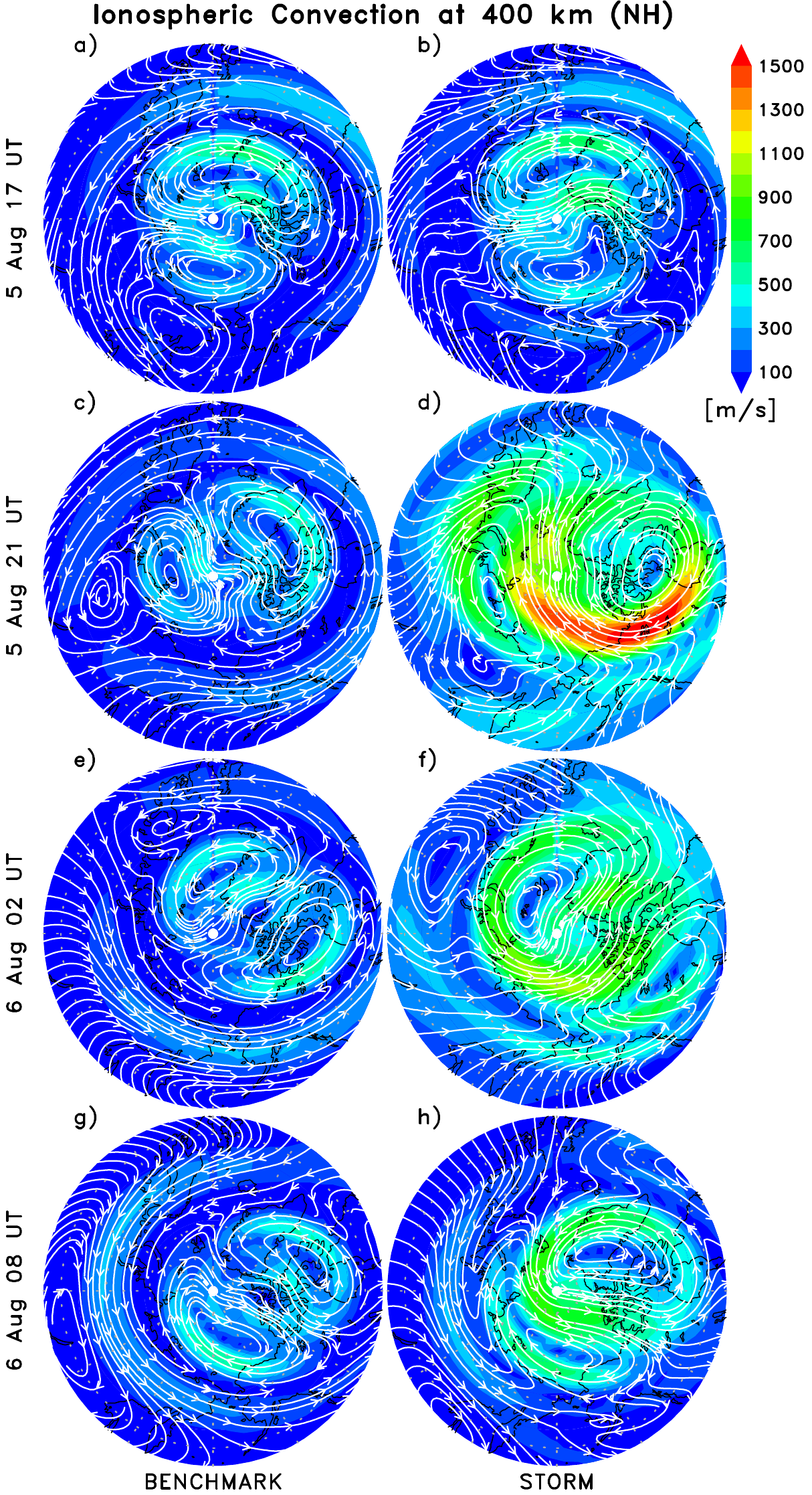}
  \caption{Northern Hemisphere ionospheric convection patterns at 400 km at four
    representative UTs on 5-6 August poleward of 51.25$^\circ$
    (geographic). Color shading is the magnitude of the horizontal ion flows in
    m s$^{-1}$. Left column is for the benchmark run, while the right column is
    for the storm run. The Greenwich Meridian is at the top of the plot. }
  \label{fig:convNH}
\end{figure}

  \subsection{High-Latitude Convection Patterns}
  We next investigate how the details of the ion convection patterns evolve
  during the storm by depicting four representative UTs. Although the
  high-latitude means of $|{\bf v_{i_{\rm H}}}|$ in the different hemispheres
  demonstrate relatively small differences in magnitude with respect to each
  other, the actual patterns of ion convection can differ greatly primarily
  because the offsets between the geographic and geomagnetic poles are different
  in the two hemispheres.

\begin{figure}[!t]\centering 
  \includegraphics[width=\columnwidth]{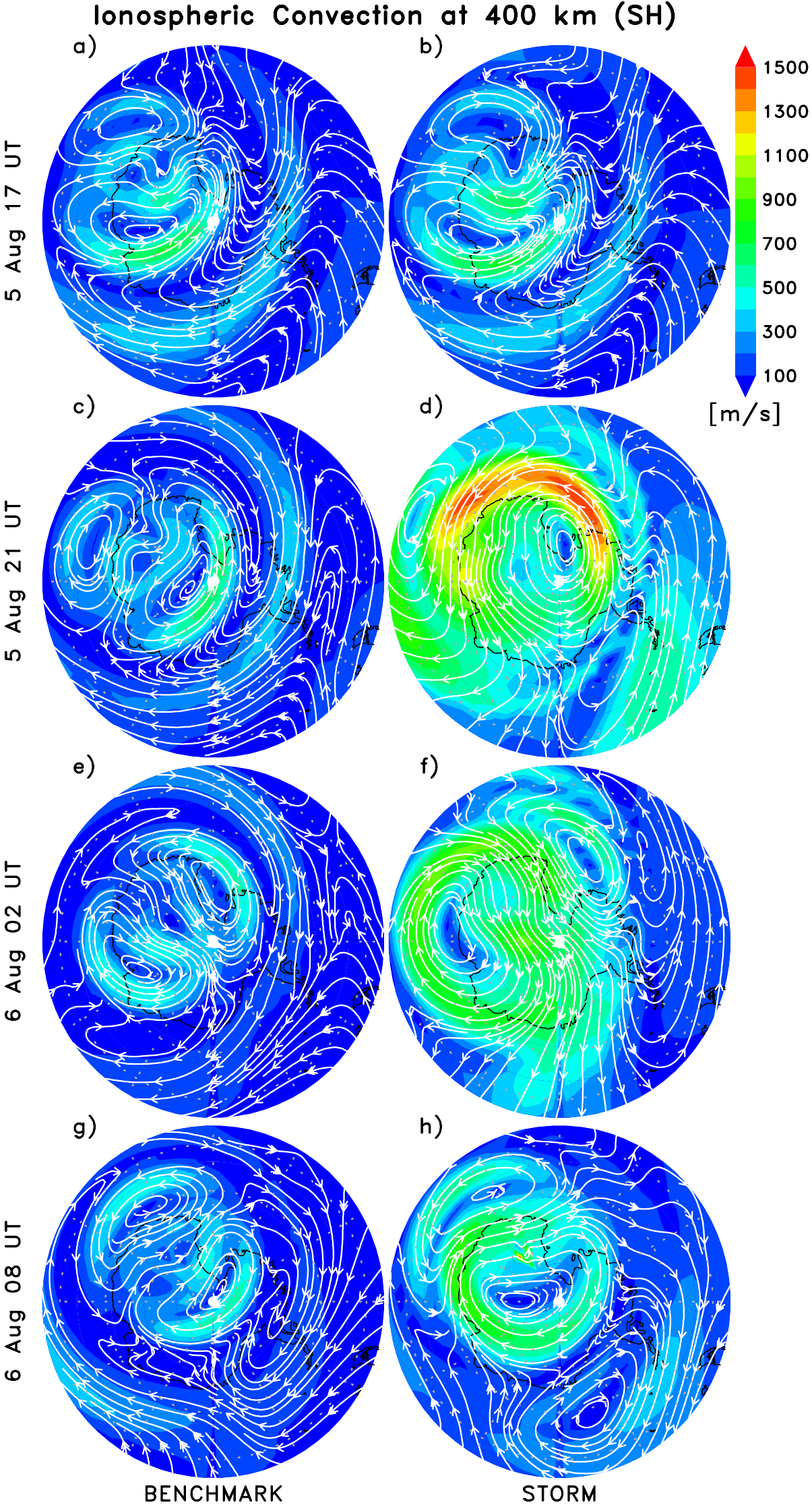}
  \caption{Same as Figure \ref{fig:convNH}, but for the Southern
      Hemisphere poleward of $-51.25^\circ$. The Greenwich Meridian is at the
      bottom of the plot.}
  \label{fig:convSH}
\end{figure}

  Figures \ref{fig:convNH} and \ref{fig:convSH} present the ionospheric
  convection patterns at 400 km in the Northern Hemisphere and Southern
  Hemispheres poleward of $\pm51.25^\circ$ (geographic), respectively. White
  streamlines represent the horizontal ion flow structures and color shading is
  the magnitude of the horizontal ion speed $|{\bf v_{i_{\rm H}}}|$. The
  benchmark run and the storm run are shown in the left and right columns,
  respectively. Four representative UTs are chosen to illustrate how the
  convection patterns evolve: (1) 1700 UT 5 August, immediately before the storm
  onset, (2) 2100 UT 5 August, main phase, (3) 0200 UT 6 August, main phase, and
  (4) 0800 UT 6 August, recovery phase.

  Before the storm onset in the Northern Hemisphere, the horizontal ion flows
  are in the storm simulation are similar to the benchmark run. This similarity
  applies both to the geographical structure and peak magnitudes of up to 500 m
  s$^{-1}$.  There is a dominant two-cell pattern at high-latitudes with a
  smaller single cell pattern at lower latitudes. Under quiet constant
  geomagnetic conditions, the pattern merely co-rotates as a function of time
  while peak flow speeds do not change much. In the storm simulation, however,
  remarkable variations are seen in the structure and magnitude of ${\bf
    v_{i_{\rm H}}}$. Ion flow speeds exceed 1600 m s$^{-1}$ during the main
  phase of the storm (panel d) within the sustained two-cell convection pattern
  that has now greatly expanded, suppressing the third cell. During the recovery
  phase of the storm maximum ion flows of up to 1100 m s$^{-1}$ are retained in
  the high-latitudes.

  Similar ionospheric response is seen in the Southern Hemisphere high-latitudes
  (Figure \ref{fig:convSH}). The dominance of the two-cell pattern is present
  under both quiet and storm-time conditions, but a significant expansion of the
  two-cell pattern is seen during the main and the recovery phases of the
  storm. Ion flow magnitudes exceed 1600 m s$^{-1}$ within the center of the
  two-cell pattern whose center is now much more offset with respect to the
  geographic South Pole.

  Remarkable response of the ion flow speeds and structure is seen in both
  hemispheres to the geomagnetic storm. However, overall, it is noteworthy that
  there exists hemispheric asymmetry during quiet periods as well as during all
  phases of the storm. We next focus on the thermospheric response to the storm.

\begin{figure*}[!t]\centering 
  \includegraphics[width=0.7\textwidth]{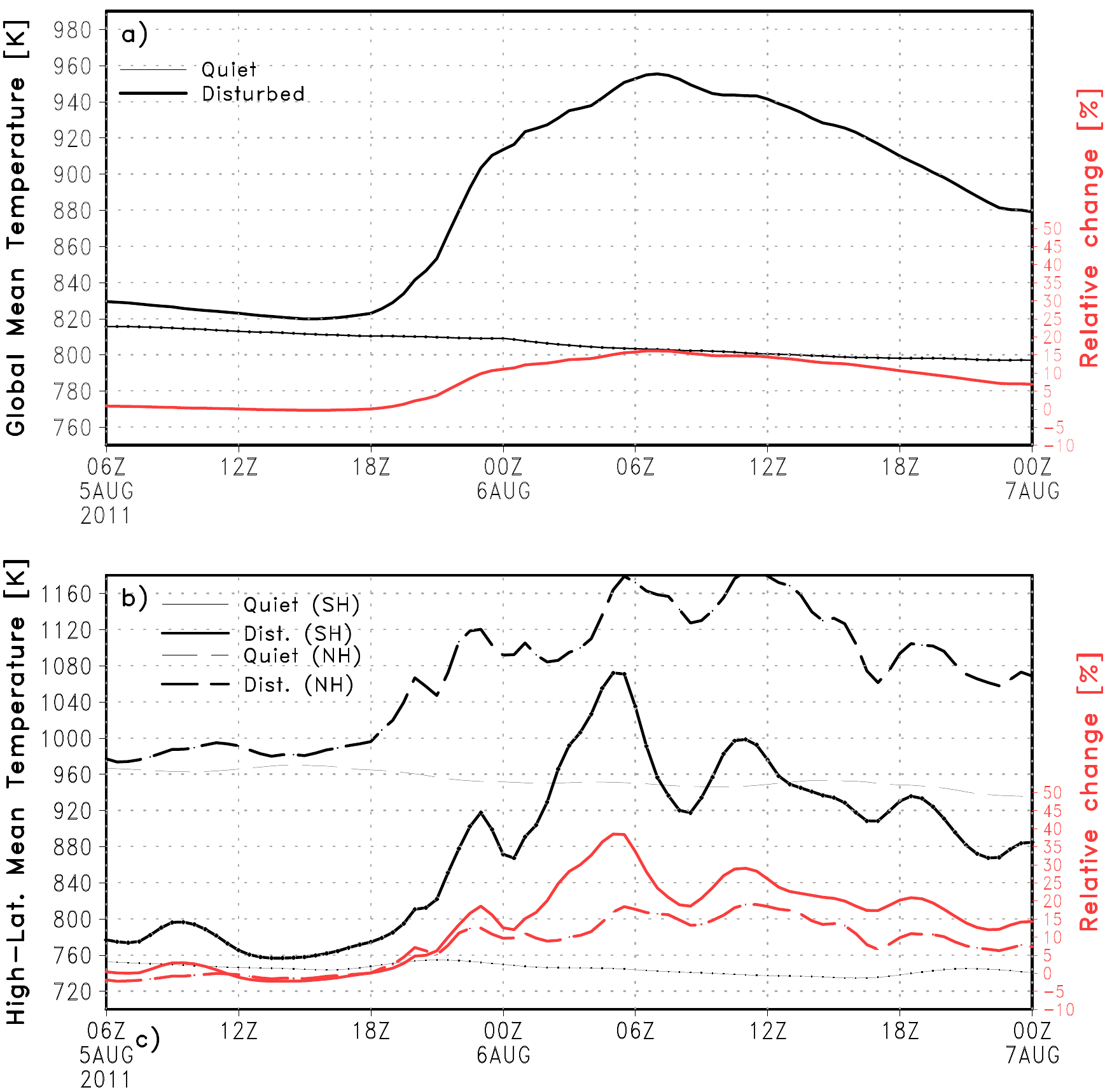}
  \caption{Variations of a) global mean and b) high-latitude mean (poleward of
    $\pm 60^\circ$) neutral temperature $T$ [K] from 5 August 0600 UT to 7
    August 0000 UT at 400 km. Benchmark and storm simulations are represented by
    thin and thick lines. In panel b, solid and dashed lines denote Southern and
    Northern Hemisphere high-latitude means, respectively. Red lines show the
    relative percentage changes with respect to the storm onset time (1800 UT, 5
    August.)}
  \label{fig:temp}
\end{figure*}

\begin{figure*}[!t]\centering 
  \includegraphics[width=0.7\textwidth]{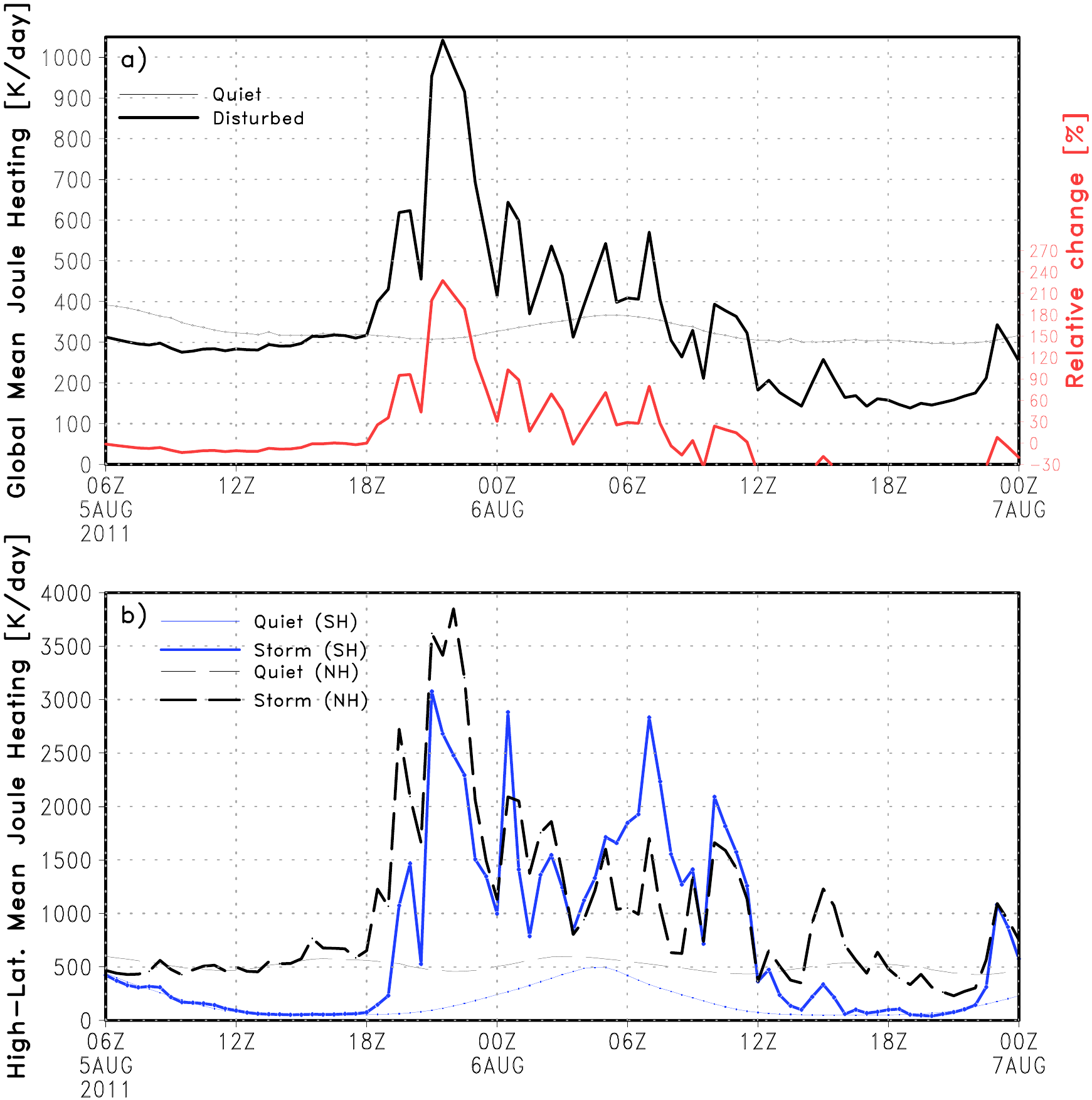}
  \caption{Variations of a) global mean and b) high-latitude mean 
    Joule heating in K day$^{-1}$ shown in the same manner as Figure
    \ref{fig:temp}.}
  \label{fig:joule}
\end{figure*}

\begin{figure*}[!t]\centering 
  \includegraphics[width=0.7\textwidth]{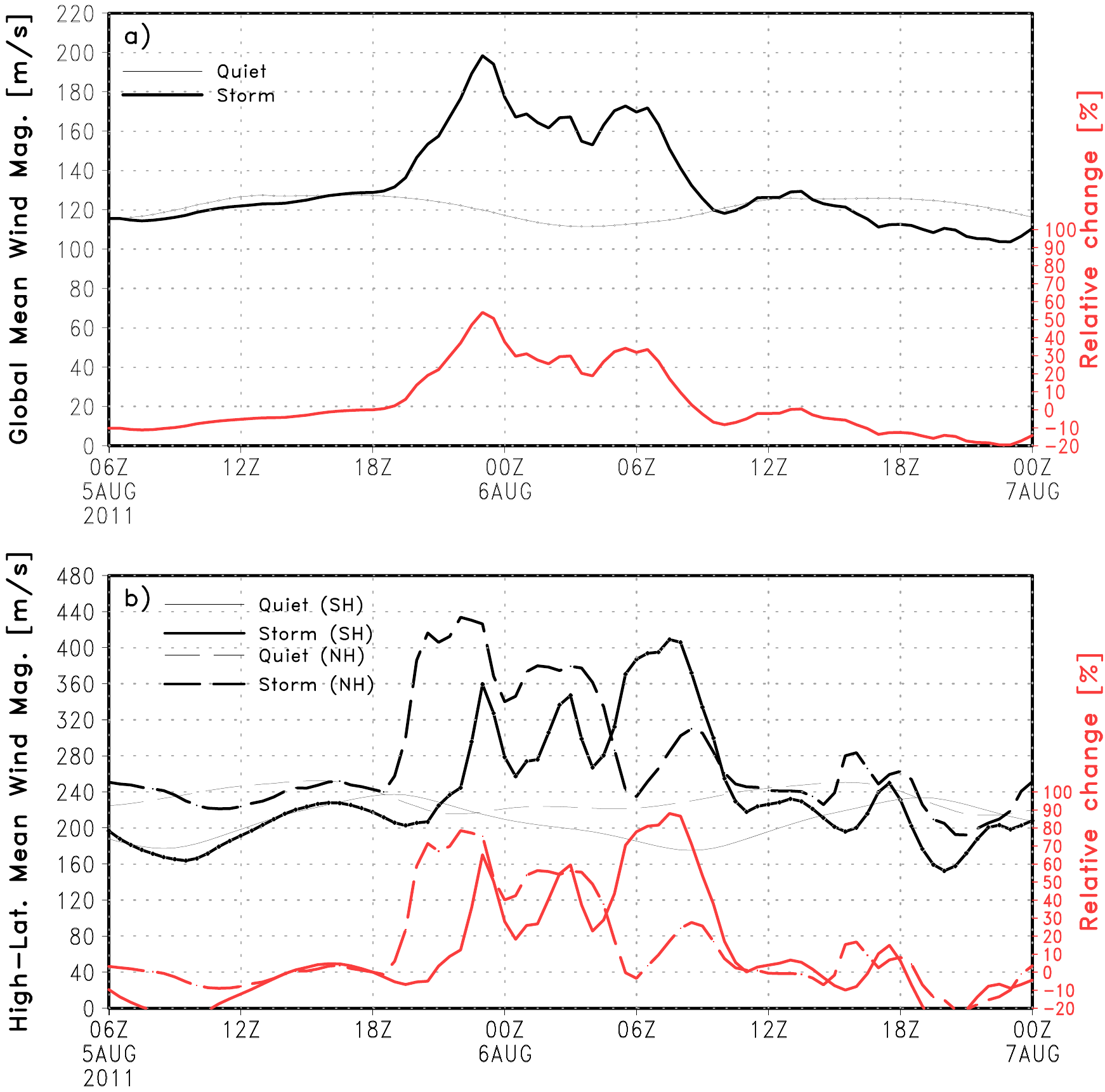}
  \caption{Same as Figure \ref{fig:temp}, but for the neutral wind magnitude.}
  \label{fig:wind}
\end{figure*}

  %
  %
  \section{Upper Atmosphere Response}
  \label{sec:thermo}
  \subsection{Global Mean Neutral Temperature and Winds}

\begin{figure}[!t]\centering\vspace{-1cm}
  \includegraphics[width=\columnwidth]{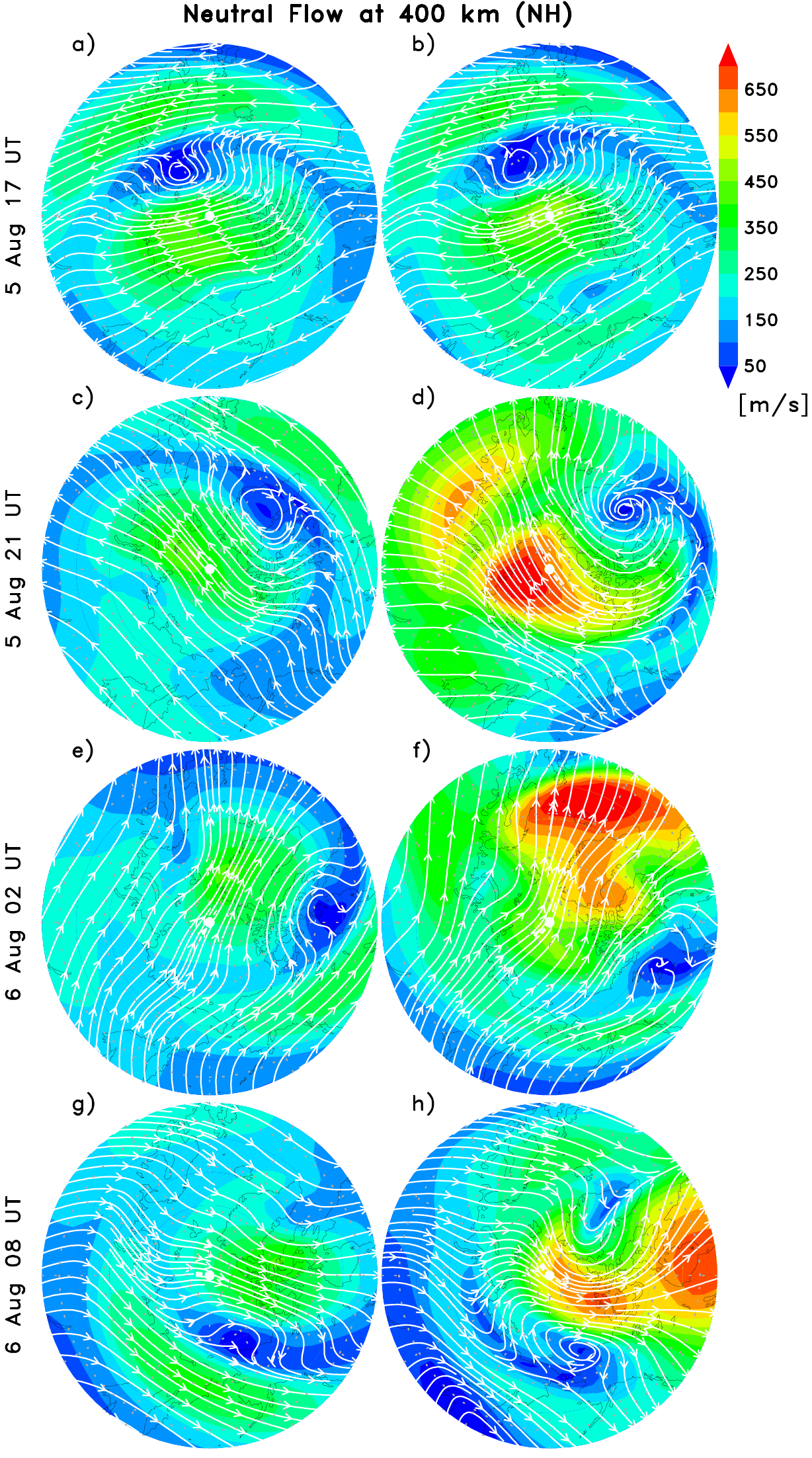}
  \caption{Same as Figure \ref{fig:convNH}, but for the Northern Hemisphere
    neutral horizontal wind flow.}
  \label{fig:uhnh}
\end{figure}

  The high-latitude ionosphere under the influence of enhanced ion convection
  can be a significant source of energy and momentum for the
  thermosphere. Figures \ref{fig:temp}a,b show how the
  global mean and high-latitude means of the neutral temperature $T$ change at
  400 km as a function of time from 5 August 0600 to 7 August 0000 UT. As the
  storm commences at around 1800 UT (5 August), the global mean temperature
  increases rapidly until 0600 UT on 6 August by 140 K ($\sim 15\%$) from 820 K
  to $\sim$960 K shown by the thick black line. In comparison, the global mean
  $T$ varies by about $\sim$10 K ($\sim 1\%$) in the same period (thin black
  line) in the quiet (benchmark) run.

  In August, the Southern and Northern Hemispheres are the winter and
  summer hemispheres, respectively, and have different dynamical and thermal
  characteristics. To quantify how the storm affects the different
  hemispheres, we evaluate the high-latitude means of neutral temperature
  poleward of $\pm 60^\circ$. Under quiet geomagnetic conditions, the summer NH
  (thin dashed line) is about 200 K warmer than the winter SH (thin solid line)
  and the mean temperature varies smoothly (diurnally) during the simulated
  period. Similarly, the storm simulation shows about 225 K temperature
  difference between the different hemispheres before the storm onset.
  
  When the storm begins, the temperatures rapidly increase in
  both hemispheres (thick lines), but the winter Southern Hemisphere peak
  thermal response to the storm (with respect to the onset time) is two times
  stronger than the summer Northern Hemisphere response: up to 40\%
  vs. $\sim$20\% situated at around 0500 UT on 6 August, 11 hours after the
  storm onset. In terms of the absolute magnitude of the thermal response, the
  peak temperature increase is $\sim$300 K in the winter Southern Hemisphere
  while it is $\sim$170 K in the summer Northern Hemisphere.

  Heating of the upper atmosphere by Joule dissipation is one of the major
  contributors to the energy budget of the high-latitude thermosphere
  \citep{Wilson_etal06}. Joule heating $Q_J$ is proportional to ionization and
  the square of the ion-neutral differential motion \citep{JohnsonHeelis05,
    YigitRidley11a}. Figure \ref{fig:joule} shows the associated neutral gas
  heating via Joule heating at 400 km in the same manner as the temperature plot
  in Figure \ref{fig:temp}. During quiet times, the global mean Joule heating
  varies smoothly between 300-400 K day$^{-1}$ over the presented period. After
  the onset of the storm the global mean Joule heating increases by more than a
  factor of three ($>200 \%$) from $\sim 300$ to more than 1000 K day$^{-1}$,
  peaking in the main phase of the storm. At the high-latitudes approximately a
  factor of six increase is seen in the summer Northern Hemisphere, and more
  than an order of magnitude increase in the winter Southern
  Hemisphere. Overall, the largest increases are seen during the initial and the
  main phase of the storm coincident with the rapid enhancements of the
  ionospheric convection that can contribute to the Joule heating
  \citep{JohnsonHeelis05}. These values suggest that instantaneously the mean
  Joule heating values at high-latitudes can exceed the mean solar heat input,
  which is typically around 1500 K day$^{-1}$.

  Similar analysis is used for the case of neutral horizontal wind magnitude
  $|{\bf u_H}| = u_H$ and we present the corresponding results in Figure
  \ref{fig:wind}. This investigation should provide an overview of how the
  mean thermospheric circulation overall responds to the storm. In the
  benchmark run (thin black line), the global mean $|{\bf u_H}|$ varies smoothly
  around 120 m s$^{-1}$. Before the storm, storm-time winds are similar to the
  benchmark run winds. However, within six hours of the storm, global mean
  horizontal wind magnitude increases $>50$ \% from $\sim$130 m s$^{-1}$ to
  $\sim$200 m s$^{-1}$. There is a secondary local peak at around 5-6 UT on 6
  August, after which the system returns gradually to quiet conditions. Overall,
  the response of the high-latitudes seen in Figure \ref{fig:wind}b is more
  rapid, intensive, and variable. Although, the wind magnitude increases rapidly
  in both hemispheres, there is a distinct difference in the timing of the peak
  response to the storm in the different hemispheres. In the NH, within the
  first three hours of the storm, $|{\bf u_H}|$ increases by 80\%, which is the
  maximum NH response during the entire storm time. On the other hand, SH
  high-latitude mean $|{\bf u_H}|$ demonstrates the largest increase at around 8
  UT on 6 August that slightly exceeds the magnitude of the peak NH response
  during the onset phase.

\begin{figure}[!t]\centering\vspace{-1cm}
  \includegraphics[width=\columnwidth]{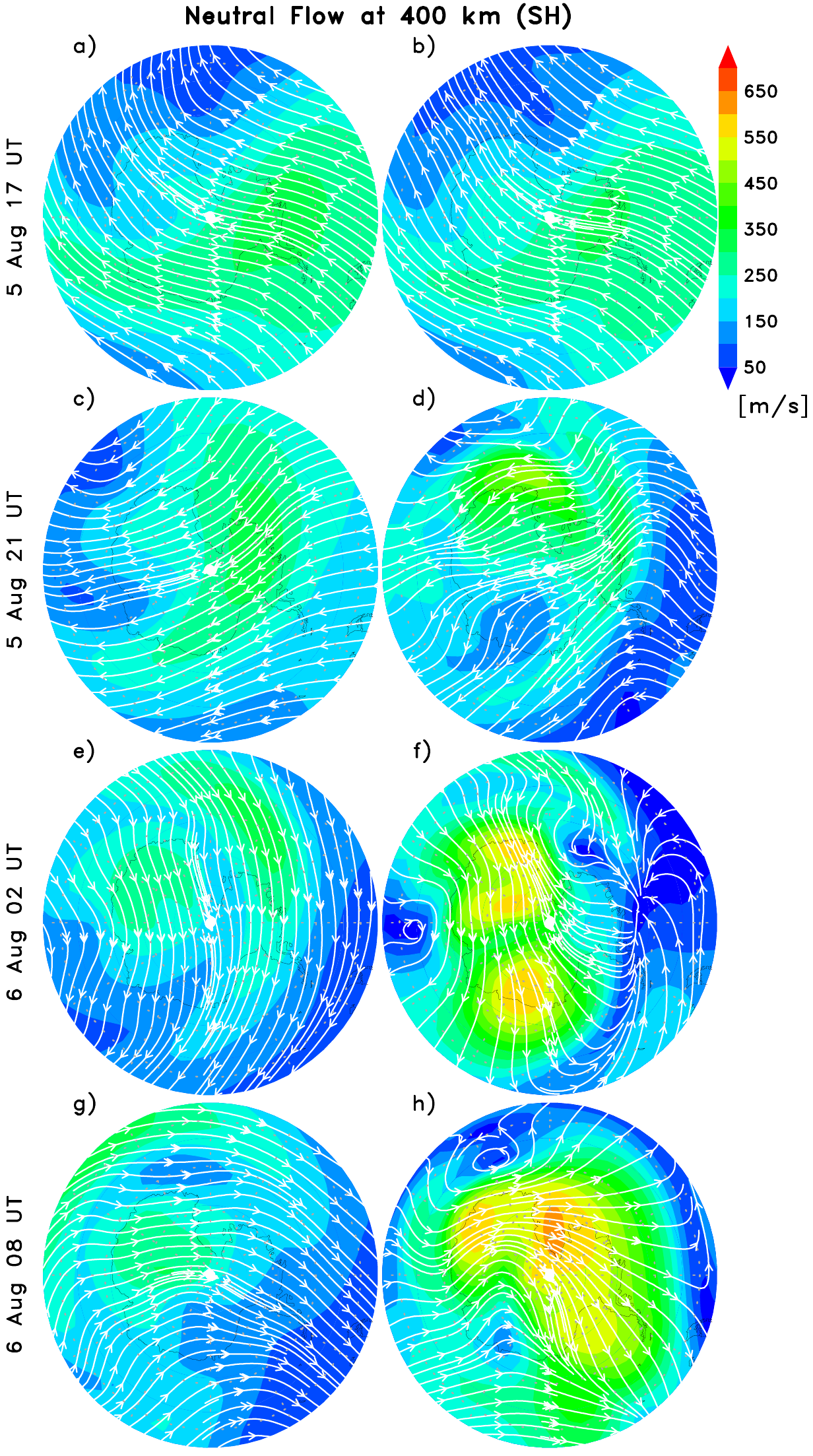}
  \caption{Same as Figure \ref{fig:convSH}, but for the Southern Hemisphere
    neutral wind flow.}
  \label{fig:uhsh}
\end{figure}

\begin{figure}[!t]\centering\vspace{-1cm}
  \includegraphics[width=\columnwidth]{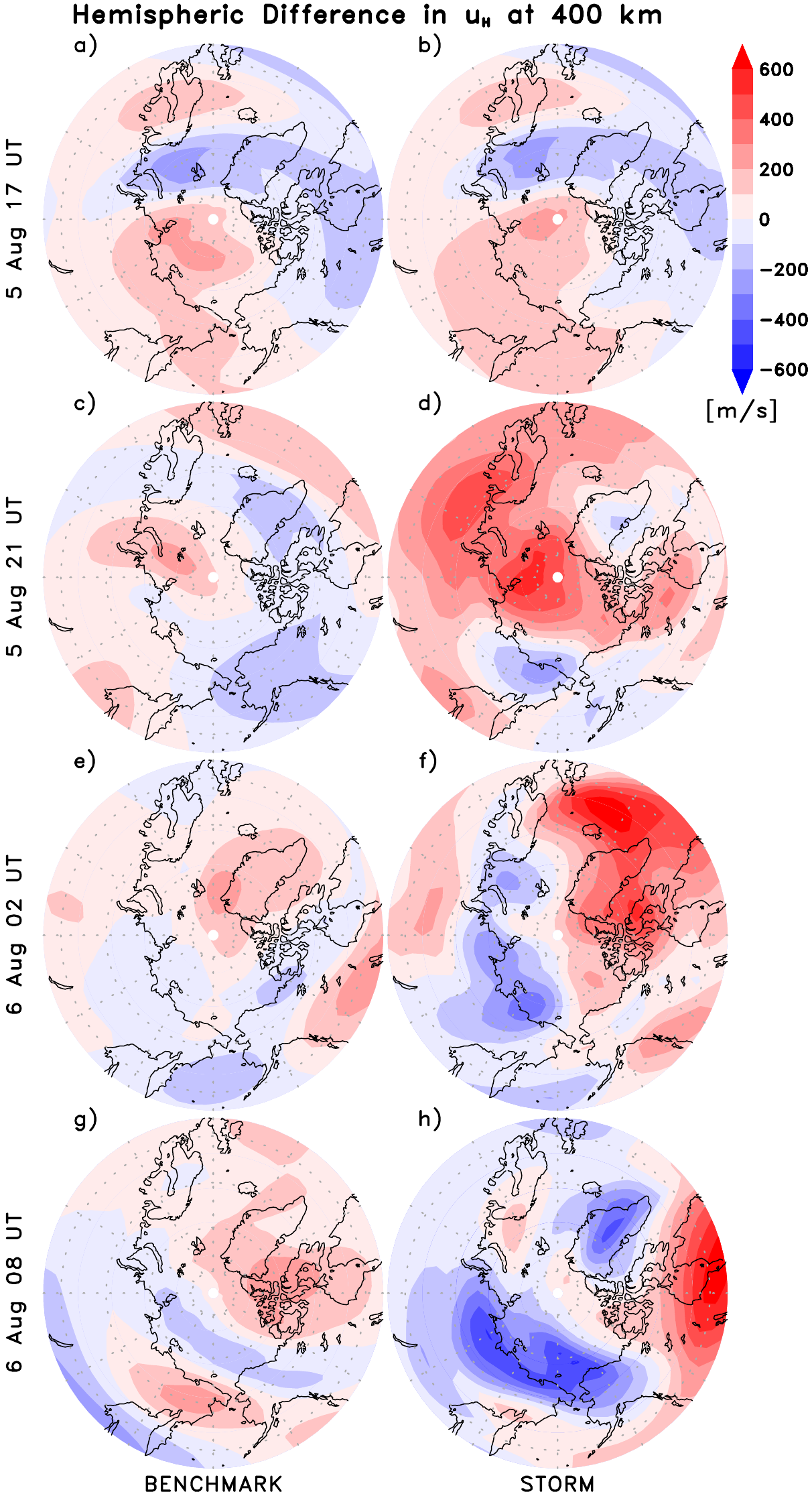}
  \caption{Hemispheric difference in the neutral horizontal wind calculated as
    the difference between the conjugate latitudes (Southern Hemisphere value
    minus Northern Hemisphere) shown poleward of $51.25^\circ$. The Northern
    Hemisphere background map is kept for reference. }
  \label{fig:hemdif}
\end{figure}

  \subsection{Effects on the Thermospheric Circulation}
  \label{sec:thermo_circulation}
  To investigate the effect of the geomagnetic storm on the upper atmosphere, we
  next study the evolution of the thermospheric circulation at high-latitudes
  during the different phases of the storm. Figures \ref{fig:uhnh} and
  \ref{fig:uhsh} present the magnitude of horizontal neutral wind $u_H$ (shaded)
  combined with the neutral flow streamlines (white) in the Northern and
  Southern Hemisphere high-latitudes, respectively, at 400 km, similar to the
  way ion convection patterns were presented in Figures
  \ref{fig:convNH} and \ref{fig:convSH}.

  In the Northern summer Hemisphere, the thermospheric circulation before the
  onset of the storm closely resembles the benchmark (quiet) simulation (Figures
  \ref{fig:uhnh}a--b) with peak $u_H$ of up to 450 m s$^{-1}$. The circulation
  pattern is characterized primarily by the injective streamlines ($u_h(t_0) \ne
  u_h(t_1)$) of a day-to-night flow that is maintained by the pressure
  difference and is modified at high-latitudes by the ion drag. There is also a
  small region of periodic motion in a region of small $u_H$ close to the
  geographic North Pole. As the storm commences, though, we can see a dramatic
  increase in the neutral flow speeds with magnitudes exceeding 700 m s$^{-1}$, while
  the benchmark simulation demonstrates a neutral circulation whose structure
  co-rotates with the local time variations and the peak flow speeds remains
  unchanged (left column, i.e., panels a, c, e, and g). In the recovery phase of
  the storm, regions of large neutral flows are still present.

  The Southern Hemisphere results shown in Figure \ref{fig:uhsh} are overall
  analogous to the Northern Hemisphere results. However, we need to note two
  important aspects. First, horizontal flows are slightly weaker in the
  winter hemisphere as the pressure forces are smaller. Also, the Southern
  Hemisphere response to the storm occurs later than the Northern
  Hemisphere. Before the storm onset, the Southern Hemisphere circulation is
  very similar to the benchmark case. Even, in the beginning of the main phase,
  the results are similar to the benchmark case. Only at the end of the main
  phase, the thermospheric circulation is enhanced in the winter hemisphere and
  can still be large during the recovery phase (panel h). This behavior is
  consistent with the high-latitude mean neutral wind speeds
  (Figure~\ref{fig:wind}).


  %
  %

  \section{Storm-Induced Hemispheric Difference in the Thermosphere} 
  \label{sec:asym}
  Storm-time evolution of the thermospheric circulation suggests that the summer
  and winter hemispheres respond differently to the storm in terms of plasma and
  neutral flow magnitudes and timing. We next investigate how the differences in
  the thermospheric circulation between the two hemispheres evolve with the
  occurrence of the storm.

  In order to evaluate the hemispheric differences in the circulation of the
  thermosphere, which we broadly term as hemispheric asymmetry, we first
  calculate the quantity $\Delta u_H$: For every longitude grid point, we
  define the hemispheric difference as the difference $\Delta f$ between the
  value of a parameter $f$ at two conjugate latitude points with respect to the
  geographic equator (i.e., Southern Hemisphere values minus Northern Hemisphere
  value). If all conjugate latitude grid points are included, then a
  latitude-longitude distribution of $\Delta f$ can be obtained.

  The results are shown in Figure \ref{fig:hemdif}. The panels are structured in
  the same way as in Figures \ref{fig:uhnh} and \ref{fig:uhsh}. We have kept the
  Northern Hemisphere background map for the purpose of convenience. Evolution
  of $\Delta u_H$ in the quiet-time run (left panels) suggest first that there
  is some degree of geographical difference between the hemispheres ($\pm 200$ m
  s$^{-1}$) in the absence of a storm, due primarily to the seasonal
  differences. Second, the amount of the difference does not change much during
  the period under consideration in the benchmark run; the structure rotates as
  a function of UT. On the other hand, during the storm $\Delta u_H$ values are
  overall much larger and exceed values of $\pm 600$ m s$^{-1}$. The peak
  magnitude of these differences is larger than the peak high-latitude mean
  horizontal wind (c.f., Figure \ref{fig:wind}b). The storm-induced rapid
  enhancement of the hemispheric difference is clearly illustrated by comparing
  $\Delta u_H$ before and after the storm onset (panels b and d). The peak
  difference values increase by a factor of 4 to 5. During the different phases
  of the storm, the magnitude as well as the structure of the hemispheric
  difference evolves in a complex manner. The difference in the dynamical
  response timing of the different hemispheres is a major factor that
  contributes to the hemispheric asymmetry at a given time.


  %
  %
  
\begin{figure*}[!ht]\centering\vspace{0cm}
  \includegraphics[width=0.7\textwidth]{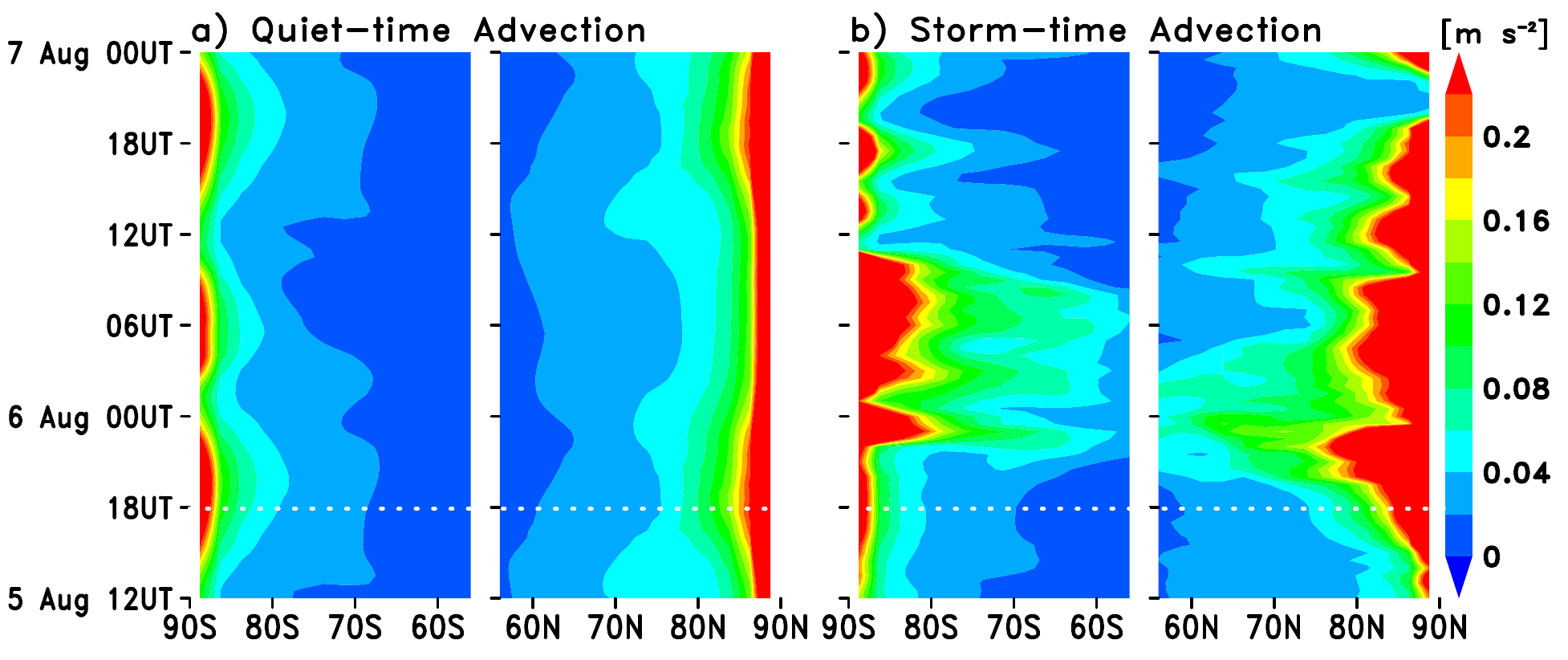}
  \caption{Temporal evolution of horizontal advective forcing at 400 km in the
    a) benchmark (left two panels) and b) the storm (right two panels)
    simulations. Units are in m s$^{-2}$. White dotted line marks the storm
    onset.}
  \label{fig:adv}
\end{figure*}
\section{Discussion}
\label{sec:discussion}
  The geomagnetic field ${\bf B}_E$ is consistently directed northward (i.e.,
  from the South Pole to the North Pole). When a strong southward interplanetary
  magnetic field (IMF) $B_{IMF}$ ($B_z<0$) arrives at the magnetosphere, the IMF
  undergoes reconnection, allowing the two field lines to connect
  temporarily. These processes can enhance energy transfer from
  the solar wind to the magnetosphere down to the thermosphere-ionosphere
  system, producing geomagnetic storms. Accordingly, our simulations show that
  the large ionospheric convection encountered in the main phase of the August
  2011 storm coincides with periods of large negative IMF ($B_z<0$).

  The asymmetric offset between the geographic and geomagnetic axes produces
  substantial differences in the momentum and heating
  sources in the upper atmosphere, greatly contributing to hemispheric
  differences in the structure and composition of the ionosphere and
  thermosphere. In August, the Northern Hemisphere is the summer hemisphere,
  while the Southern Hemisphere is in winter. Therefore, a certain degree of
  difference will be present because of the seasonal (solar irradiation)
  differences between the hemispheres when the geomagnetic storm occurs.
  
  In our simulations, remarkable dynamical changes are seen in the thermospheric
  circulation, following the rapid changes in the ionospheric
  convection due to the enhanced electric fields of magnetospheric
  origin, which influence the ion drift motion ${\bf v_i}$ given
  approximately by
  \begin{linenomath*}
    \begin{equation}
      \label{eq:1}
      {\bf v}_{{\bf E}\times {\bf B}} = \frac{{\bf E} \times {\bf B}}{|{\bf B}|^2}.
    \end{equation}
  \end{linenomath*} 
  Overall, the structure of the simulated ion convection patterns are consistent
  with the previously observed \citep[e.g.,][]{Bristow_etal11} and modeled
  \citep{KilleenRoble84, HeppnerMaynard87} patterns at high-latitudes. During
  $B_{z}<0$ conditions, the large-scale structure of this convection is
  characterized by anti-sunward flow over the polar cap and return flow at
  auroral latitudes.
  
  Changes in the general circulation of the thermosphere occur during
  the storm due primarily to the variations in the pressure
  and ion drag forcings. These are two major dynamical mechanisms that shape the
  neutral flow. During a geomagnetic storm the neutral temperature is enhanced
  which modulates the pressure force. On the other hand, enhanced ion flows
  (Figures \ref{fig:conv}--\ref{fig:convSH}) can drive the neutrals to higher
  speeds via increased ion drag effect.  Storm-time increased conductivities and
  electric fields lead to an increase in the current density. From the point of
  thermospheric dynamics, ion motion is viewed in terms of drift
  velocities. Therefore the momentum exchange between the ions and neutrals is
  proportional to their differential velocity ${\bf v_i}-{\bf u}$. At
  high-latitudes, ions possess larger speeds and can thus impart an ion drag
  force on the neutrals. Increased neutral flows can then produce additional
  nonlinear dynamical response in the system in the later stages of the storm
  because of advective processes.

  Based on the above dynamical analysis, the structure and the evolution of the
  simulated hemispheric differences in the thermospheric circulation can be
  interpreted qualitatively as the following. In the earlier stages of the
  storm, enhanced ion flows drive the neutrals to larger speeds, to a different
  degree in the different hemispheres. By the beginning of the recovery phase of
  the storm, the thermospheric circulation has become stronger and therefore,
  nonlinear processes in the neutral flow, such as wind advection, sustain large
  winds and contribute to hemispheric differences. The associated horizontal
  wind advection is shown in Figure \ref{fig:adv}, where the UT-latitude
  cross-sections of the horizontal advective forcing of the neutral horizontal
  circulation are shown for the quiet-time (panel a) and the storm-time
  (panel b) simulations, respectively. The advective forcing is enhanced
  appreciable during the storm and is characterized by a marked hemispheric
  asymmetry in terms of its spatiotemporal variations. In this context, our
  simulations support the conclusions of \citet{Forster_etal08} that a non-zero
  $B_y$ is likely to contribute to an asymmetry in the thermosphere. During the
  storm, substantial fluctuations are seen in $B_y$ as well. Recently,
  \citet{Yamazaki_etal15} have highlighted the importance of the $B_y$ effect in
  structure of the thermosphere.

  The models of
  \citet{Fuller-RowellRees80} and \citet{Dickinson_etal81} were the first of
  their kind in terms of three-dimensional modeling of the thermosphere. They
  have used simplified ionospheres represented by the \citet{Chiu75} empirical
  ionosphere model and the configuration of Earth's geomagnetic field has been
  simplified by a dipole field. One of the first studies of thermospheric
  circulation during geomagnetic storms was conducted by
  \citet{Fuller-RowellRees81} for idealized conditions of a substorm specified
  by the variation of the $K_p$ index from 1 to 6. Their simulation covered a
  period of 4.5 h. Wind speeds exceeding 600 m s$^{-1}$~ were seen at around 400
  km. In our study we have used the measured storm-time IMF values from the ACE
  satellite in order to specify the configuration of the high-latitude electric
  fields.

  Recently, the works by \citet{Haaser_etal13}, \citet{Earle_etal13},
  \citet{Gong_etal13}, and \citet{Huang_etal14} have studied the low-latitude
  dynamics during the August 2011 storm using global satellite observations and
  incoherent scatter radar (ISR) data. \citet{Gong_etal13}'s analysis of Arecibo
  ISR data showed profound meridional wind enhancements in the low-latitude
  thermosphere. Also, \citet{Earle_etal13} showed, using the Air Force
  Communication/Navigation Outage Forecasting System (C/NOFS) satellite, an
  increase in the neutral horizontal flows during the storm. Our simulations
  have shown that the global mean horizontal neutral winds are overall enhanced
  (up to 50 \%) during the storm and this increase is even more pronounced in
  the high-latitudes (up to 80\%). Our results are consistent
  with these observations.

  \section{Summary and Conclusions}
  \label{sec:conc}
  
  Using the three-dimensional nonhydrostatic Global
  Ionosphere Thermosphere Model (GITM) driven by the observed solar and
  geomagnetic activity input from satellites, dynamical and thermal response of
  the thermosphere and ionosphere to the 5--6 August 2011 major geomagnetic
  storm has been quantified and new physical insights to the challenging problem
  of hemispheric differences have been provided.
  We have analyzed the major ion and neutral parameters during the
  storm. Specifically, high-latitude mean ion and neutral flows, global mean
  temperature, polar stereographic projections of the
  ion and neutral flows in the Northern and Southern Hemispheres have been
  evaluated during the different phases of the storm and are compared to
  quiet-time simulations. The magnitude of hemispheric differences in the
    thermospheric circulation has been determined and interpreted by calculating
    nonlinear advective forcing. Polar stereographic projections have been
  performed for representative storm periods: Immediately before the storm
  onset, main storm phase, and the recovery phase.
  The main findings are: 
  
  (1) Storm-induced changes in the ionosphere show rapid
  and large enhancement of ion flows with the onset of the storm and the
  simulated ion convection patterns are consistent with previous modeling and
  observational studies;
  
  (2) Thermospheric circulation changes are appreciable during the storm. The
  response of the neutrals to the storm is a slower process, owing to their
  larger inertia. The global mean neutral temperature increases by up to
  $15\%$. At high-latitudes, the thermal response of the Southern winter
  Hemisphere is overall two times larger than the Northern summer Hemisphere:
  $30\%$ vs. $15\%$. The global mean neutral wind magnitude increases up to
  $50\%$ and up to $80\%$ in the high-latitude mean winds. The response of the
  neutral winds in the winter Southern Hemisphere occurs later than in the
  summer Northern Hemisphere; and

  (3) Substantial hemispheric differences are seen during the storm in the
  thermospheric circulation resulting from ion-neutral coupling effects and
  nonlinear dynamical changes. Comparison with quiet-time simulations
  suggest that storm-time hemispheric differences are at least a factor of two
  larger and highly variable during the different phases of the storm.

  (4) Especially in the recovery phase of the storm, a significant degree of
    hemispheric difference is seen in the global circulation of the
    thermosphere. Modeled nonlinear advective processes in the neutral
    horizontal flows demonstrate substantial differences between the two
    hemispheres, suggesting that advective forcing plays an important role in
    maintaining hemispheric differences in the thermosphere.

    Here, we have not included the effects of lower atmospheric small-scale
    waves on the upper atmosphere as the model lower boundary is around the
    mesopause. Gravity waves propagate to the upper atmosphere
    \citep{Yigit_etal08, GavrilovKshevetskii13} and produce appreciable
    dynamical \citep{Yigit_etal09, Yigit_etal12b} and thermal effects
    \citep{YigitMedvedev09} on the general circulation of the thermosphere up to
    F-region altitudes. Thermospheric gravity wave effects exhibit substantial
    solar cycle variations \citep{YigitMedvedev10}. Under moderate major
    geomagnetic storm conditions, the dynamical effects of internal waves on the
    upper atmosphere are probably of minor significance at the altitudes ($\sim
    400$ km) studied in this paper. However, a detailed quantification of
    internal wave effects on the upper atmosphere during geomagnetic storms is
    yet to be done. Global models extending into the upper atmosphere
    \citep[e.g.,][]{Yigit_etal09} or in general, whole atmosphere models
    \citep[e.g.,][]{Jin_etal11, Akmaev11} could be used to study internal
    wave effects at higher altitudes in conjunction with space
    weather effects.

  One limitation of our simulations is the use of the empirical model of
  \citet{Weimer96}. Empirical models provide a gross (mean) structure of the
  atmosphere and do not include the realistic variability of the upper
  atmosphere. Empirical ionospheric convection models, specifically, provide an
  average structure of high-latitude convection patterns, based on a large
  collection of observations \citep{Bekerat_etal03}. The principle is similar to
  the distribution of neutral winds modeled by empirical wind models
  \citep[e.g.,][]{Hedin_etal96} or the solar irradiance models
  \citep{Tobiska_etal00}. Empirical models are broadly used in the aeronomy
  community due to their simplicity and portability.  

  In future investigations, assimilative modeling techniques could provide a
  better ground for representing (small-scale) variability at high-latitudes
  during disturbed geomagnetic conditions.  A comparison of convection patterns
  obtained from the Assimilative Mapping of Ionospheric Electrodynamics
  technique \citep[AMIE,][]{RichmondKamide88} with patterns retrieved from the
  DMSP satellite showed that AMIE patterns matched the observations better than
  statistical models \citep{Bekerat_etal05}. Future research on storm-induced
  thermospheric variations could utilize the AMIE technique.

\noindent

\section*{Acknowledgements}
\label{sec:acknowledgements}

We have used the Disturbed Storm Time index ($Dst$) provided by the World Data
Center for Geomagnetism, Kyoto.  Erdal Yi\u{g}it was partially supported by NASA
grant NNX13AO36G and George Mason University's tenure-track faculty summer
fellowship.


\newpage

\end{document}